\documentstyle[epsf,12pt]{article}
\newcommand{\Section}[1]{\setcounter{equation}{0} \section{#1}}
\newcommand{\rf}[1]{(\ref{#1})}
\newcommand{\beq}[1]{ \begin{equation}\label{#1} }
\newcommand{\eeq}{\end{equation} }
\newcommand{\ba}{\begin{array}}
\newcommand{\ea}{\end{array}}
\newcommand{\beqa}{\begin{eqnarray}}
\newcommand{\eeqa}{\end{eqnarray}}

\newtheorem{definition}{Definition}[section]

\newcommand{\hff}{{\scriptstyle \frac{1}{2}}}
\newcommand{\hf}{{\textstyle \frac{1}{2}}}

\newcommand{\LF}{L}
\newcommand{\derc}{r}
\begin{document}
\bibliographystyle{plain}
\hfill {\footnotesize Lecture Notes for short course, AMS-93}

\bigskip
\begin{center}
\begin{large}
{\bf Wavelets and Fast Numerical Algorithms}
\end{large}

\bigskip

G. Beylkin \\
Program in Applied Mathematics \\
University of Colorado at Boulder \\
Boulder, CO 80309-0526

\end{center}

Numerical algorithms using wavelet bases are similar to other transform methods in that vectors and operators are expanded into a basis and the computations take place in the new system of coordinates. As in all transform methods, such approach seeks an advantage in that the computation is faster in the new system of coordinates than in the original domain. However, due to the recursive definition of wavelets, their controllable localization in both space and wave number (time and frequency) domains, and the vanishing moments property, wavelet based algorithms exhibit a number of new and important properties.
 
In the usual transform methods, the functions of the basis (e.g. exponentials, Chebyshev polynomials, etc.) are chosen to be eigenfunctions of some differential operator (e.g. solutions of the Sturm-Liouville problem). The choice of the differential operator and, hence, of the basis functions, is dictated by the availability of fast algorithms for expanding an arbitrary function into the basis. Unfortunately, classes of operators which have a sparse representation in such bases are very narrow. 

Wavelets, on the other hand, are not solutions of a differential equation. These functions are defined recursively and are generated via an iterative algorithm. They are translations and dilations of a single function 
\footnote{It is possible to construct bases with translations and dilations of several functions, see e.g. \cite{ABCR}.}.
Instead of diagonalizing some differential operator, representations in the wavelet bases reduce a wide class of operators to a sparse form. Here the orthogonality of wavelets to the low degree polynomials (the vanishing moments property) plays a crucial role \footnote{This property and the fact that the basis is orthonormal, distinguish the wavelet bases from the hierarchical bases.}.

The orthonormal bases of wavelets were fisrt constructed by Stromberg \cite{STROMBERG} and then by Meyer \cite{MEYER0}. Later, the notion of the Multiresolution Analysis was introduced by Meyer \cite{MEYER86} and Mallat \cite{MALLAT}. Orthonormal bases of compactly supported wavelets were constructed by Daubechies \cite{DAUB1}. There are many new constructions of orthonormal bases with a controllable localization in the time--frequency domain, notably "wavelet-packet" bases in \cite{CO-MEY89} and \cite{CO-WICK90}, local trigonometric bases in \cite{CO-MEY90} and \cite{MALVAR90}, wavelet bases on the interval in \cite{DAUB-COHEN-JA-VI}, \cite{COHEN-DAUB-VI} and \cite{JOUINI-LR}. Very important connection exists between the wavelets and the technique of subband coding in signal processing. In fact, the discrete wavelet transform is accomplished by the pair of the so-called quadrature mirror filters. The exact quadrature mirror filters (QMFs) were introduced by Smith and Barnwell \cite{SMITH-BARN}.

Wavelets have some of their historical roots in Littlewood-Paley and Calder\'on-Zygmund theories (see e.g. \cite{MEYER}) which has been powerful tools in analysis of linear and non-linear operators. In Numerical Analysis some of the ingredients of Calder\'on-Zygmund theory appear in the Fast Multipole Method (FMM) for computing potential interactions \cite{ROKHLIN1}, \cite{GR-RO}, \cite{CA-GR-RO}. FMM was designed for computing potential interactions between $N$ particles in $O(-\log \epsilon N)$ operations (instead of $O(N^2)$ operations). The reduction of the complexity in FMM is achieved by approximating  the far field effect of a cloud of charges located in a box  by the effect of a single multipole at the center of the box. All boxes are then organized in a dyadic hierarchy enabling an efficient $O(N)$ algorithm.

Fast wavelet-based algorithms of \cite{BCR1} provide a systematic generalization of the FMM and its descendents (e.g. \cite{DONNEL-RO}, \cite{ALPERT-RO}, \cite{GREENGARD90}) to all Calder\'on-Zygmund and pseudo-differential operators. The subdivision of the space and its organization in a dyadic hierarchy are a consequence of the multiresolution properties of the wavelet bases, while the vanishing moments of the basis functions make them useful tools for approximation.

A novel aspect of representing operators in the wavelet bases is the so-called non-standard form \cite{BCR1}. The remarkable feature of the non-standard form is the uncoupling of the interactions between the scales. The non-standard form leads to an order $N$ algorithm for evaluating operators on functions. It is also quite remarkable that the error estimates for the non-standard form  lead to a proof of the selebrated \lq\lq T(1)" theorem of David and Journ\'e (see \cite{BCR1}). The non-standard forms of many basic operators, such as derivatives, fractional derivatives, the Hilbert and Riesz transforms,  may be computed explicitly \cite{BEY}. 
A straightforward realization, or the standard form, by contrast, contains matrix entries reflecting \lq\lq interactions" between all pairs of scales. The standard form yields, in general, only an order $ N \log(N) $ algorithm for evaluating operators on functions. 
 
The representation of wide classes of operators in wavelet bases may be viewed as a method for their \lq\lq compression", i.e., conversion to a sparse form. For these operators sparse representations lead to fast algorithms for matrix multiplications. Since the performance of many algorithms requiring multiplication of dense matrices has been limited by $O(N^3)$ operations, these fast algorithms address a critical numerical issue.

Among the algorithms requiring multiplication of matrices is an iterative algorithm for constructing the generalized inverse \cite{SCHULZ}, the scaling and squaring method for computing the exponential of an operator, and similar algorithms for sine and cosine of an operator, to mention a few. By replacing the ordinary matrix multiplication in these algorithms by the fast multiplication in the wavelet bases, the number of operations is reduced to, essentially,  an order $N$ operations. For example, if both, the operator and its generalized inverse, admit sparse representations in the wavelet basis, then the iterative algorithm \cite{SCHULZ} for computing the generalized inverse requires only $O(N \log \kappa)$ operations, where $\kappa$ is the condition number of the matrix. Various numerical examples and applications may be found in \cite{BCR2}, \cite{ABCR} and \cite{BCRREV}.

Solving the two-point boundary value problem for the elliptic differential operators in the wavelet \lq\lq system of coordinates" allows us to construct the Green's function (the inverse operator) in $O(N)$ operation.  We note that the ordinary matrix representation of the Green's function requires $O(N^2)$ significant entries but the representation of the Green's function in the wavelet bases requires (for a given accuracy) only $O(N)$ entries. The main tool in constructing the Green's function numerically is the diagonal preconditioner available for the periodized differential operators in the wavelet bases \cite{BEY}, \cite{BEYBVP} (see also \cite{JAFFARD}).

Unfortunately, the format of one lecture does not allow us to cover all the developments or mention all the results available today. Instead, we will review several features of the new numerical methodology based on the wavelet representations. Starting from the notion of multiresolution analysis, we will consider the non-standard form (which achieves decoupling among the scales) and the associated fast numerical algorithms. Examples of non-standard forms of several basic operators (e.g. derivatives) will be computed explicitly.

\Section{Multiresolution Analysis and Wavelets.}
\label{MRAandWAVE}

\bigskip

We briefly outline here the properties of compactly supported wavelets and refer for details to \cite{DAUB1}, \cite{DAUB10} and \cite{MEYER}. 
Let us start with the notion of the multiresolution analysis \cite{MEYER86}, \cite{MALLAT} which captures the essential features of all multiresolution approaches developed so far. 
\begin{definition}
\label{MRA}

Multiresolution analysis  is a 
decomposition of the Hilbert space ${\bf L}^2 ({\bf R}^{\bf d})$,
${\bf d} \ge 1$, into a chain of closed subspaces
\beq{chain}
\dots \subset {\bf V}_2 \subset {\bf V}_1 
\subset {\bf V}_0 \subset {\bf V}_{-1} \subset {\bf V}_{-2} \subset \dots
\eeq
such that
\begin{enumerate}
\item 
$\bigcap_{j \in {\bf Z}} {\bf V}_j = \{ 0 \}$ and
$\bigcup_{j \in {\bf Z}} {\bf V}_j$ is  dense in 
${\bf L}^2 ({\bf R}^{\bf d})$

\item For any $f \in {\bf L}^2 ({\bf R}^{\bf d})$ and any $j \in {\bf Z}$,
$f(x) \in {\bf V}_j $ if and only if $f(2 x) \in {\bf V}_{j-1}$

\item For any $f \in {\bf L}^2 ({\bf R}^{\bf d})$ 
and any $k \in {\bf Z}^{\bf d}$,
$f(x) \in {\bf V}_0 $ if and only if $f(x-k) \in {\bf V}_0$

\item There exists a scaling function $\varphi \in {\bf V}_0$ such that
$\{ \varphi(x-k) \}_{k \in {\bf Z}^{\bf d}}$ is a Riesz basis of ${\bf V}_0$.

\end{enumerate}
\end{definition}

\noindent In this lecture we use only orthonormal bases, so that we replace
Condition~4 by

\bigskip

{\em

\noindent $ \ \ \ \ $ 4'. There exists a scaling function $\varphi \in {\bf V}_0$ such that $\{ \varphi(x-k) \}_{k \in {\bf Z}^{\bf d}}$ is an orthonormal basis of ${\bf V}_0$.

}
\bigskip

Let us define the subspaces ${\bf W}_j$ as an orthogonal complement of ${\bf V}_j$ in ${\bf V}_{j-1}$,
\beq{1.0001}
{\bf V}_{j-1} = {\bf V}_j \oplus {\bf W}_j,
\eeq
and represent the space ${\bf L}^2 ({\bf R}^{\bf d})$ as a direct sum
\beq{1.0002}
{\bf L}^2 ({\bf R}^{\bf d}) = \bigoplus_{j \in {\bf Z}} {\bf W}_j.
\eeq
Selecting the coarsest scale $n$, we may replace the chain of the subspaces
\rf{chain} by
\beq{chain0}
{\bf V}_n \subset \dots \subset {\bf V}_2 
\subset {\bf V}_1 \subset {\bf V}_0 \subset {\bf V}_{-1} 
\subset {\bf V}_{-2} \subset \dots,
\eeq
and obtain 
\beq{1.00020}
{\bf L}^2 ({\bf R}^{\bf d}) 
= {\bf V}_n \bigoplus_{j \le n} {\bf W}_j.
\eeq
If there is a finite number of scales then, without 
loss of generality, we set $j=0$ to be the finest scale and consider 
\beq{chain1}
{\bf V}_n \subset \dots \subset {\bf V}_2 \subset {\bf V}_1 \subset {\bf V}_0 , 
\quad {\bf V}_0  \subset {\bf L}^2 ({\bf R}^{\bf d})
\eeq
instead of \rf{chain0}. In numerical realizations the subspace ${\bf V}_0$ is finite dimensional.

The function $\varphi$ is the so-called scaling function and, with its help, we  may define the function $\psi$, the wavelet, such that the set of functions  $\{ \psi(x-k) \}_{k \in {\bf Z}}$ is an orthonormal basis of ${\bf W}_0$, 

An example of the multiresolution analysis satisfying Definition~\ref{MRA} with Condition~4' is the chain of subspaces generated by the Haar basis \cite{HAAR}.
The scaling function in this case is the characteristic function of the interval $(0,1)$. The Haar function is defined as
\beq{v01}
h(x)= \left \{
\ba{ll}
1 & {\rm for} \;\; 0 < x < 1/2  \\
-1 & {\rm for} \;\;  1/2 \leq x < 1 \\
0 & {\rm elsewhere}.
\ea
\right. ,
\eeq
and the Haar basis is formed by functions  $h_{j,k}(x)=2^{-j/2}h(2^{-j} x-k)$ $j,k \in {\bf Z} $.

The wavelet bases (with a smooth scaling function $\varphi$ of Condition~4') generalizing the Haar functions were first constructed by Stromberg \cite{STROMBERG} and then Meyer \cite{MEYER0}. The notion of the  Multiresolution Analysis was introduced by Meyer \cite{MEYER86} and Mallat \cite{MALLAT} and it is more recent than the constructions of \cite{STROMBERG}, \cite{MEYER0} and, of course, of \cite{HAAR}. Compactly supported wavelets with vanishing moments were constructed by I.~Daubechies \cite{DAUB1} and we will use them in this lecture. However, most of the results that we discuss do not depend on this particular choice of the wavelet bases.

The vanishing moments property simply means that the basis functions are chosen to be orthogonal to the low degree polynomials, namely, if the set of functions  $\{ \psi(x-k) \}_{k \in {\bf Z}}$ is an orthonormal basis of ${\bf W}_0$, then
\beq{1.019}
\int_{-\infty}^{+\infty} \psi(x) x^m dx=0, \qquad  m=0,\dots,M-1.
\eeq
For the Haar function in \rf{v01} $M=1$ and it is trivially orthogonal to constants.

There are two immediate consequences of Definition~\ref{MRA} with Condition~4'.
First, the function $\varphi$ may be expressed as a linear combination of the basis functions of ${\bf V}_{-1}$. Since the functions 
$\{\varphi_{j,k}(x)= 2^{-j/2} \varphi (2^{-j} x-k)\}_{k \in {\bf Z}}$
form an orthonormal basis of  ${\bf V}_j$,  we have
\beq{1.001}
\varphi (x) = \sqrt{2} \sum_{k=0}^{\LF - 1}  h_{k} \varphi (2x-k).
\eeq
In general, the sum in \rf{1.001} does not have to be finite and, by choosing the finite sum in \rf{1.001}, we are selecting the compactly supported wavelets. We may rewrite \rf{1.001} as
\beq{1.002}
{\hat \varphi} (\xi) = m_0 (\xi/2) {\hat \varphi} (\xi/2),
\eeq
where
\beq{1.003}
{\hat \varphi} (\xi) = \frac{1}{\sqrt{2 \pi}}
\int_{-\infty}^{+\infty}  \varphi (x) \, {\rm e}^{{\rm i} x \xi} \, dx,
\eeq
and
the $2 \pi$-periodic function  $m_0$ is defined as
\beq{1.004}
m_0(\xi)  = 2^{-1/2} \sum_{k=0}^{\LF - 1} h_k {\rm e}^{{\rm i} k \xi}.
\eeq

Second, the orthogonality of $\{ \varphi(x-k) \}_{k \in {\bf Z}}$ implies that
\beq{1.005}
\delta_{k0} = 
\int_{-\infty}^{+\infty} \varphi(x-k) \varphi(x) \, dx =
\int_{-\infty}^{+\infty} | {\hat \varphi} (\xi) |^2 \,
{\rm e}^{-{\rm i} k \xi} \, d\xi,
\eeq
and, therefore,
\beq{1.006}
\delta_{k0} =
\int_{0}^{2\pi}  \sum_{l \in {\bf Z}} |{\hat \varphi} (\xi+2 \pi l) |^2 \,
{\rm e}^{-{\rm i} k \xi} \, d\xi,
\eeq
and
\beq{1.007}
\sum_{l \in {\bf Z}} |{\hat \varphi} (\xi+2 \pi l) |^2  = 1.
\eeq
Using \rf{1.002}, we obtain 
\beq{1.008}
\sum_{l \in {\bf Z}} 
|m_0 (\xi/2 + \pi l)|^2 |{\hat \varphi} (\xi/2 +\pi l) |^2  = 1,
\eeq
and, by taking the sum in \rf{1.008} separately over odd and even indeces, we have
\beq{1.008a}
\sum_{l \in {\bf Z}} 
|m_0 (\xi/2 + 2 \pi l)|^2 |{\hat \varphi} (\xi/2 + 2 \pi l) |^2 +
\sum_{l \in {\bf Z}} 
|m_0 (\xi/2 + 2 \pi l + \pi )|^2 |{\hat \varphi} (\xi/2 + 2 \pi l + \pi) |^2  = 1.
\eeq
Using the $2 \pi$-periodicity of the function  $m_0$ and \rf{1.007}, we obtain
(after replacing  $\xi/2$ by $\xi$) a necessary condition 
\beq{1.009}
| m_0(\xi)|^2 + | m_0(\xi+ \pi)|^2 = 1,
\eeq
for the coefficients $h_k$ in \rf{1.004}. On denoting
\beq{1.014}
m_1 (\xi) = {\rm e}^{-{\rm i} \xi} {\overline m_0} (\xi+\pi),
\eeq
and defining the function $\psi$,
\beq{1.017}
\psi (x) = \sqrt{2} \sum_{k} g_{k} \varphi (2x-k),
\eeq
where
\beq{1.016}
g_k=(-1)^{k} h_{\LF - k-1},\qquad k=0,\dots,\LF-1,
\eeq
or, the Fourier transform of $\psi$,
\beq{1.018}
{\hat \psi} (\xi) = m_1 (\xi/2) {\hat \varphi} (\xi/2),
\eeq
where
\beq{1.015}
m_1 (\xi) = 2^{-1/2} \sum_{k=0}^{k=\LF - 1} g_k {\rm e}^{{\rm i} k \xi},
\eeq
it is not difficult to show (see e.g., \cite{MEYER}, \cite{DAUB1}, \cite{DAUB10}), that on each fixed scale $j \in {\bf Z}$, the wavelets 
$\{\psi_{j,k}(x) = 2^{-j/2} \psi (2^{-j} x-k)\}_{k \in {\bf Z}}$ 
form an orthonormal basis of  ${\bf W}_j$.

Equation \rf{1.009} defines a pair of the quadrature mirror filters (QMFs) $H$ and $G$, where $H=\{h_k\}_{k=0}^{k=\LF - 1}$ and $G=\{g_k\}_{k=0}^{k=\LF - 1}$. The exact QMF filters were first introduced by Smith and Barnwell \cite{SMITH-BARN} for subband coding.

We will not go into the details of considering necessary and sufficient conditions for the quadrature mirror filters $H$ and $G$ to generate the wavelet basis and refer to \cite{DAUB10} for the details. The coefficients of the quadrature mirror filters $H$ and $G$ are computed by solving a set of algebraic equations (see e.g. \cite{DAUB10}). The number  $\LF$ of the filter coefficients in \rf{1.004} and \rf{1.015} is related to the number of vanishing moments $M$, and $\LF=2M$ for the wavelets constructed in \cite{DAUB1}. If additional conditions are imposed (see \cite{BCR1} for an example), then the relation might be different, but $\LF$ is always even.

We observe that once the filter $H$ has been chosen, it completely determines the functions $\varphi$ and $\psi$ and therefore, the multiresolution analysis. Moreover, in properly constructed  algorithms, the values of the functions  $\varphi$ and $\psi$ are (almost) never computed. Due to the recursive definition of the wavelet bases, all the manipulations are performed with the quadrature mirror filters $H$ and $G$, even if they involve quantities associated with $\varphi$ and $\psi$. 

As an example, let us compute the moments of the scaling function $\phi$.
The expressions for the moments, 
\beq{1.030}
{\cal M}^m_\infty = \int x^m \, \varphi (x) \ dx, \quad m=0,\dots,M-1,
\eeq
may be  found in terms of the filter coefficients $\{h_k\}_{k=1}^{k=\LF}$ using 
\beq{1.031}
{\hat \varphi} (\xi) = (2\pi)^{-1/2} 
\prod_{j=1}^{\infty} \ m_0(2^{-j}\xi),
\eeq
where $m_0(\xi)$ is given in \rf{1.004}.

The moments  ${\cal M}^m_\infty $ are obtained (within the desired accuracy) by recursively generating a sequence of vectors,
$\{ {\cal M}^m_r \}_{m=0}^{m=M-1}$ for  $r=1,2,\dots,$
\beq{1.033}
{\cal M}^m_{r+1}= \sum_{j=0}^{j=m}  
\left( \ba{l} m \\ j \ea \right) 2^{-jr}
{\cal M}^{m-j}_{r} {\cal M}^j_1 ,
\eeq
starting with 
\beq{1.034}
{\cal M}^m_{1}= 2^{- m - \frac{1}{2}} 
\sum_{k=0}^{k=\LF-1}  h_k k^m, \quad m=0,\dots,M-1.
\eeq
Each vector $\{ {\cal M}^m_r \}_{m=0}^{m=M-1}$ represents  $M$ moments of 
the product in \rf{1.031} with  $r$ terms, and the iteration converges rapidly. Notice, that we never computed the values of the function $\varphi$ itself. 

%
%
%
\Section{The non-standard form}
\label{NSF}

The wavelet bases in ${\bf L}^2 ({\bf R}^{\bf d})$, $d \ge 2$, may be constructed as a tensor product of the one-dimensional bases. Considering $d=2$ and using the Haar basis as an example, we note that the supports of the basis functions are rectangles of various dyadic sizes. Representing operators in such bases leads to the standard form which we will discuss in the next Section. 

Alternatively, wavelet bases in ${\bf L}^2 ({\bf R}^{\bf d})$, $d \ge 2$ may be constructed using the scaling function in addition to the wavelets. Such construction is specific to wavelet bases. Considering $d=2$ as an example, we note that the triplet of functions
\beq{2dbasis}
\{ \psi_{j,k}(x) \, \psi_{j,k'}(y), \ \psi_{j,k}(x)  \, \varphi_{j,k'}(y) , \ \varphi_{j,k}(x)  \,  \psi_{j,k'}(y) \},
\eeq
where $j, k, k'  \in {\bf Z}$, forms the basis of  ${\bf L}^2 ({\bf R}^{\bf 2})$. We note that the basis functions have square supports.  Representing operators in these bases leads to the non-standard form \cite{BCR1}. 

Let us introduce the non-standard form in the context of the Multiresolution Analysis, independently of the specific choice of the wavelet basis.
Let $T$ be an operator
\beq{10.r1}
T : {\bf L}^2 ({\bf R}) \to {\bf L}^2 ({\bf R}),
\eeq
with the kernel $K(x,y)$.
We define projection operators on the subspace ${\bf V}_j$, $j \in {\bf Z}$,
\beq{10.r2}
P_j : {\bf L}^2 ({\bf R}) \to {\bf V}_j,
\eeq
as follows
\beq{v11}
\left( P_jf \right) (x) =\sum_{k} \langle f, \varphi_{j,k} \rangle \varphi_{j,k} (x).
\eeq
Expanding
$T$ in a \lq\lq telescopic" series, we obtain
\beq{1.r10}
T = \sum_{j \in {\bf Z}} (Q_jTQ_j + Q_jTP_j + P_jTQ_j),
\eeq
where
\beq{v14}
Q_{j}=P_{j-1}-P_{j}
\eeq
is the projection operator on the subspace ${\bf W}_j$.
If there is the coarsest scale $n$, then instead of \rf{1.r10} we have
\beq{1.r11}
T = \sum_{j=-\infty}^{n} (Q_jTQ_j + Q_jTP_j + P_jTQ_j) +P_nTP_n,
\eeq
and if the scale $j=0$ is the finest scale, then
\beq{1.r12}
T_0 = \sum_{j=1}^{n} (Q_jTQ_j + Q_jTP_j + P_jTQ_j)+P_nTP_n,
\eeq
where $T  \sim T_0=P_0 T P_0$ is a discretization of the operator $T$ on the finest scale.

The non-standard form is a representation (see \cite{BCR1})
of the operator $T$ as a chain of triplets
\beq{10.22}
T=\{ A_j, B_j, \Gamma_j \}_{j \in {\bf Z}}
\eeq
acting on the subspaces ${\bf V}_j$ and ${\bf W}_j$,
\beq{10.23}
A_j : {\bf W}_j \to {\bf W}_j,
\eeq
\beq{10.24}
B_j : {\bf V}_j \to {\bf W}_j,
\eeq
\beq{10.25}
\Gamma_j : {\bf W}_j \to {\bf V}_j.
\eeq
The operators $\{ A_j, B_j, \Gamma_j \}_{j \in {\bf Z}}$ are defined as 
$A_j = Q_j T Q_j$, $B_j = Q_j T P_j$ and $\Gamma_j  = P_j T Q_j$.
These operators admit a recursive definition via the relation 
\beq{10.26}
T_j = 
\left(\begin{array}{cc}
A_{j+1} & B_{j+1} \\
\Gamma_{j+1}  & T_{j+1} \end{array} \right),
\eeq
where operators $T_j = P_j T P_j$,
\beq{10.27}
T_j : {\bf V}_j \to {\bf V}_j.
\eeq

If there is a coarsest scale $n$, then 
\beq{10.29}
T=\{ \{ A_j, B_j, \Gamma_j \}_{j \in {\bf Z} : j \le n}, T_n\},
\eeq
where $T_n= P_n T P_n$.
If the number of scales is finite, then $j=1,2,\dots,n$ in \rf{10.29} and
the operators are organized as blocks of the matrix (see Figures~1~and~2).
\begin{figure}
\centerline{\epsffile{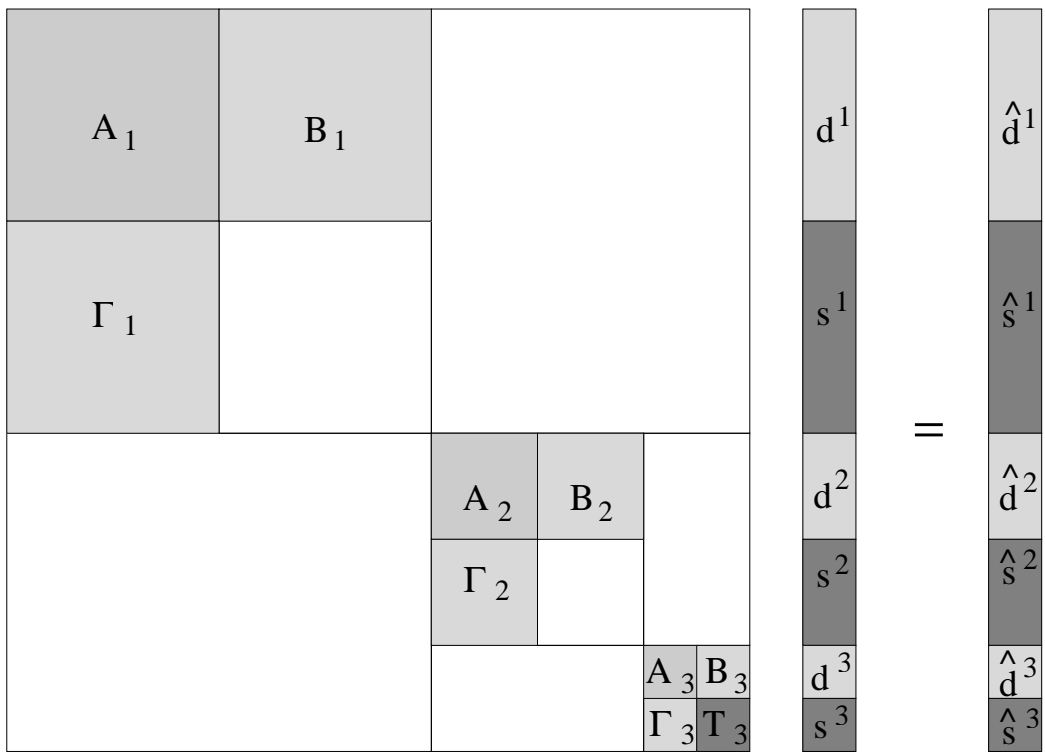}}
\caption{Organization of the non-standard form of a matrix. The submatrices
$A_j$, $B_j$, and $\Gamma_j$, $j=1,2,3$, and $T_3$ are the only non-zero
submatrices. }
\label{Fig1}
\end{figure}
\begin{figure}
\epsfxsize = 300pt
\centerline{\epsffile{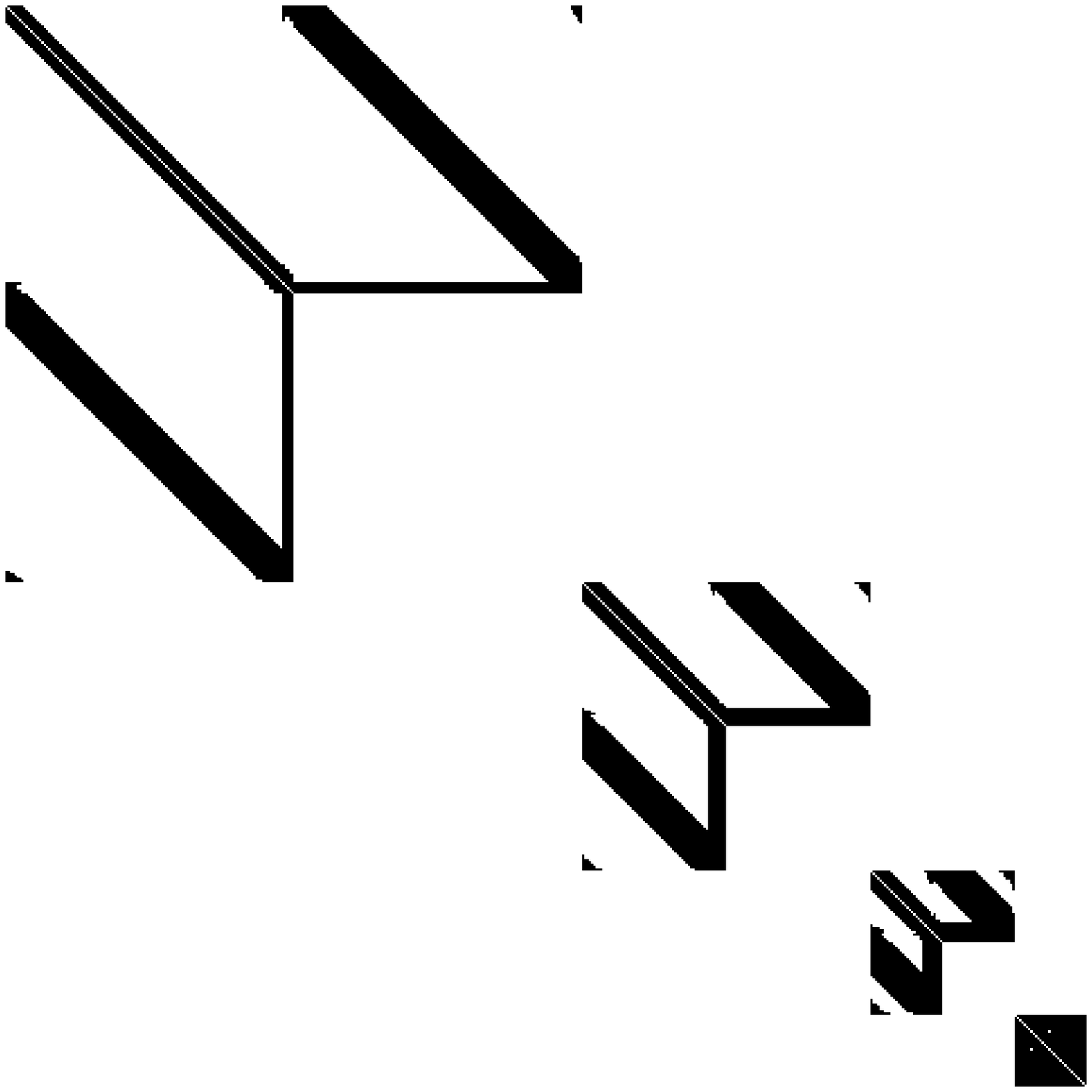}}
\caption{An example of a matrix in the non-standard form (see Example~1)}
\label{Fig2}
\end{figure}
\begin{figure}
\epsfxsize = 300pt
\centerline{\epsffile{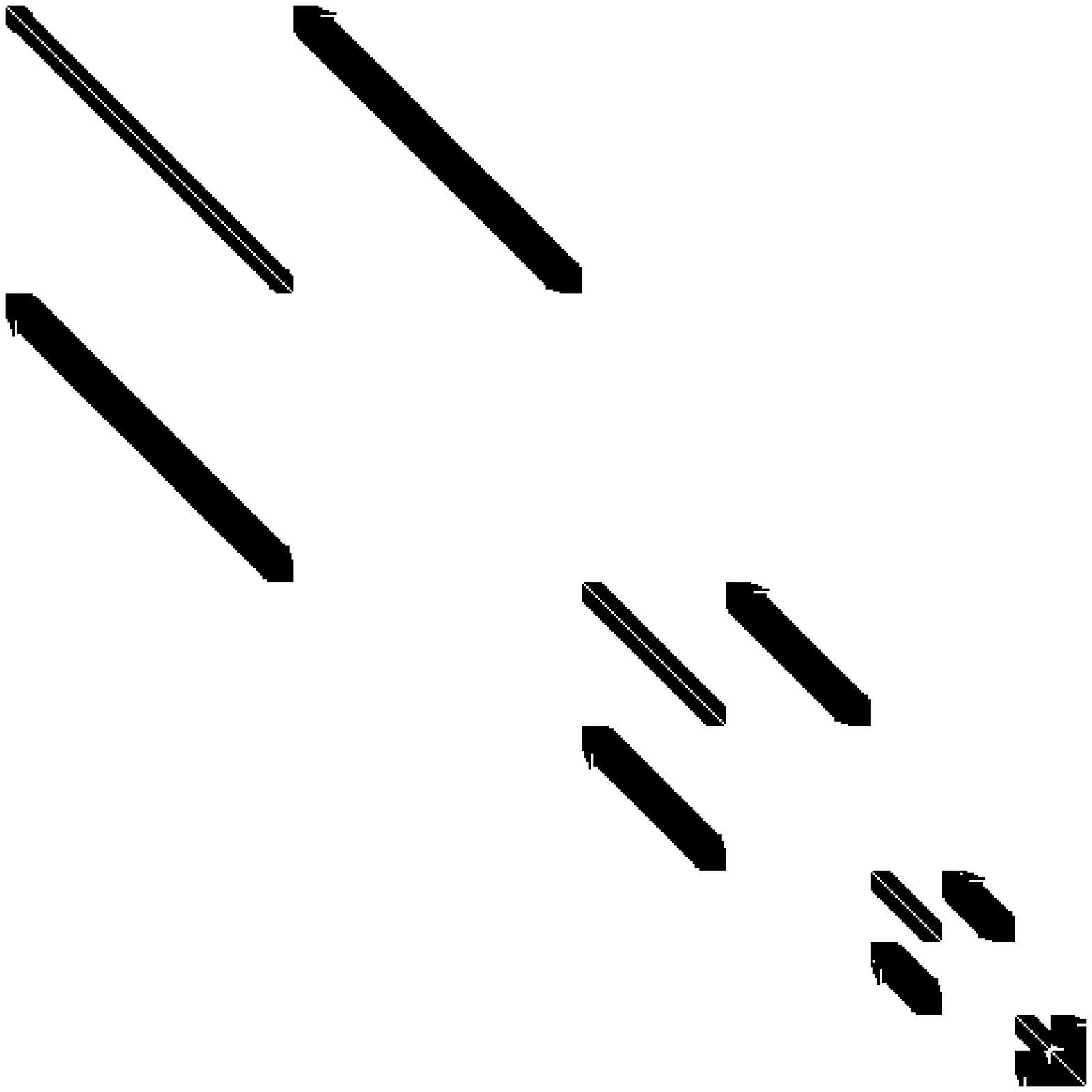}}
\caption{The non-standard form of the same matrix as in  Figure~\protect\ref{Fig2} in the basis of the wavelets on the interval \protect\cite{BEY-BREW}. The vertical and horizontal bands (which are present in Figure~\protect\ref{Fig2} due to periodization) do not appear in this representation}
\label{Fig2a}
\end{figure}
Let us make the following observations:

\noindent 1). The map \rf{10.23} implies that the operator $A_j$ 
describes the interaction on the scale $j$ only, since the subspace
${\bf W}_j$ is an element of the direct sum in \rf{1.00020}.

\noindent 2). The operators $B_j$, $\Gamma_j$ in \rf{10.24} and \rf{10.25}
describe the interaction between scale
$j$ and all coarser scales.  
Indeed, the subspace ${\bf V}_j$ contains all the subspaces
${\bf V}_{j'}$ with $j' > j$ (see \rf{chain}).

\noindent 3). The operator $T_j$ is an \lq\lq averaged" version of the operator $T_{j-1}$.

The operators  $ A_j$, $ B_j$ and $\Gamma_j$ 
are represented by  the matrices 
$\alpha^j$, $\beta^j$ and $\gamma^j$, 
\beq{1.34a}
\alpha^j_{k, k^\prime} = \int \int K(x,y) \, \psi_{j,k} (x) 
\, \psi_{j,k^\prime} (y) \ dx dy,
\eeq
\beq{1.34b}
\beta^j_{k, k^\prime} = \int \int K(x,y) \, \psi_{j,k} (x) \,
\varphi_{j,k^\prime} (y) \ dx dy,
\eeq
and
\beq{1.34c}
\gamma^j_{k, k^\prime} = \int \int K(x,y) \, \varphi_{j,k} (x) \, 
\psi_{j,k^\prime} (y) \ dx dy.
\eeq
The operator $T_j$ is represented by the matrix $s^j$,
\beq{1.34s}
s^j_{k, k^\prime} = \int \int K(x,y) \, \varphi_{j,k} (x) \, 
\varphi_{j,k^\prime} (y) \ dx dy.
\eeq
%
%
%
%
%
%


\Section{The standard form}
\label{SF}

The standard form is the representation of an operator in the tensor product basis. Instead of introducing the standard form in this manner, we emphasize the connection with the non-standard form. The standard form is obtained by representing 
\beq{10.30}
{\bf V}_j = \bigoplus_{j' > j} {\bf W}_{j'},
\eeq
and considering for each scale $j$ the
operators $\{ B_j^{j'}, \Gamma_j^{j'} \}_{j' > j}$,
\beq{10.31}
B_j^{j'} : {\bf W}_{j'} \to {\bf W}_j,
\eeq
\beq{10.32}
\Gamma_j^{j'} : {\bf W}_j \to {\bf W}_{j'}.
\eeq
If there is the coarsest scale $n$, then instead of \rf{10.30} we have
\beq{10.33}
{\bf V}_j = {\bf V}_n \bigoplus_{j'=j+1}^{j'=n} {\bf W}_{j'}.
\eeq
In this case, the operators $\{ B_j^{j'}, \Gamma_j^{j'} \}$  for $j'=j+1,\dots,n$  
are as in \rf{10.31} and \rf{10.32} and,
in addition, for each scale
$j$ there are operators $\{ B_j^{n+1}\}$ and $\{ \Gamma_j^{n+1} \}$,
\beq{10.34}
B_j^{n+1} : {\bf V}_{n} \to {\bf W}_j,
\eeq
\beq{10.35}
\Gamma_j^{n+1} : {\bf W}_j \to {\bf V}_{n}.
\eeq
(In this notation, $\Gamma_n^{n+1} = \Gamma_n$ and $B_n^{n+1} = B_n$).
If there are finitely many scales and ${\bf V}_0$ is finite dimensional, then 
the standard form is a representation
of $T_0=P_0 T P_0$ as 
\beq{10.36}
T_0=\{ A_j, \{ B_j^{j'} \}_{j'=j+1}^{j'=n}, \{ \Gamma_j^{j'} \}_{j'=j+1}^{j'=n},
B_j^{n+1}, \Gamma_j^{n+1}, T_n\}_{j=1,\dots,n}.
\eeq
The operators  \rf{10.36} are organized as blocks of the matrix (see Figures~3~and~4). 

\begin{figure}
\centerline{\epsffile{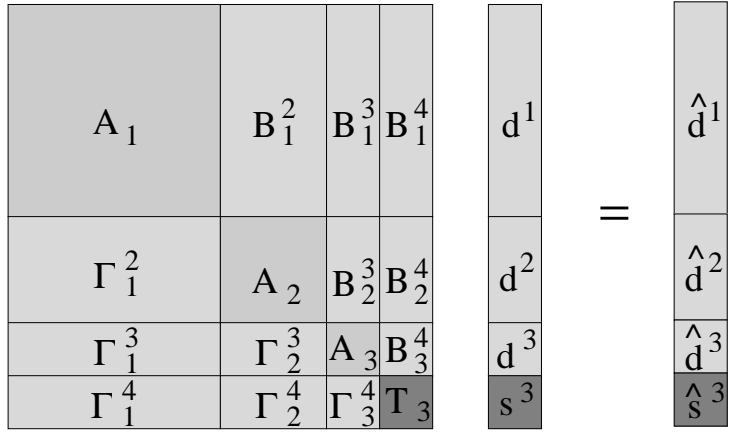}}
\caption{Organization of a matrix in the standard form}
\label{Fig3}
\end{figure}

\begin{figure}
\epsfxsize = 225pt
\centerline{\epsffile{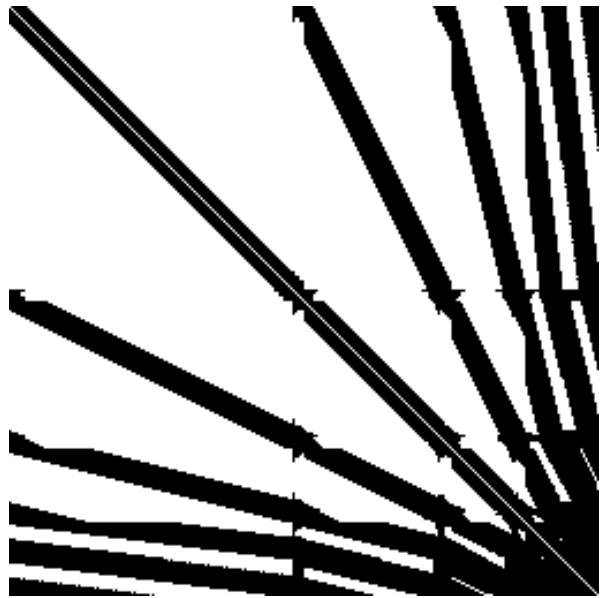}}
\caption{An example of a matrix in the standard form (see Example~1)}
\label{Fig4}
\end{figure}

If the operator  $T$ is a 
Calder\'on-Zygmund or a pseudo-differential operator
then, for a fixed accuracy, all the operators in \rf{10.36} (except $T_n$) are banded.
As a result, the standard form has  several \lq\lq finger" 
bands which correspond to the interaction between different scales.
For a large class of operators 
(pseudo-differential, for example), the interaction between different scales
characterized by the size of the coefficients of \lq\lq finger" 
bands, decays
as the distance $j'-j$ between the scales increases. Therefore, if the scales
$j$ and $j'$ are well separated, then
for a given accuracy, 
the operators $B_j^{j'}, \Gamma_j^{j'}$ can be neglected.  

There are two ways of computing the standard form of a matrix. First consists
in applying the one-dimensional transform to each
column (row) of the matrix and, then, to each row (column) of the result.
Alternatively, one can compute the non-standard form and then
apply the one-dimensional transform to each row of all operators $B^j$ and
each column of all operators $ \Gamma_j$. We refer to \cite{BCR1} for details.

\Section{Compression of operators}
\label{Comp}

If the operator
$T$ is a Calderon-Zygmund or a pseudo-differential operator then, by
using the wavelet basis with $M$ vanishing moments, we force operators
$\{ A_j, B_j, \Gamma_j \}_{j \in {\bf Z}}$ to decay roughly as $1/d^{M+1}$, where 
$d$ is a distance from the diagonal. 
For example, let the kernel satisfy the conditions
\beqa
| K(x,y)| & \leq & \frac{1}{| x-y|},  \label{3.00} \\
|\partial_x^M K(x,y)|+|\partial_y^M K(x,y)|
& \leq & \frac{C_0}{| x-y|^{1+M}}
\label{3.01}
\eeqa
for some $M \geq 1$.
Then by choosing the wavelet basis with $M$ vanishing moments,
the coefficients
$ \alpha_{i,l}^j, \beta_{i,l}^j, \gamma_{i,l}^j $
of the non-standard form (see \rf{1.34a} - \rf{1.34c})
satisfy the estimate
\beq{3.02}
|\alpha^j_{i,l}|+|\beta^j_{i,l}| +|\gamma^j_{i,l}|
\leq\frac{C_M}{1+  | i-l|^{M+1}},
\eeq
 for  all
\beq{3.02x}
 \mid i-l \mid  \geq 2 M.
\eeq
If, in addition to \rf{3.00}, \rf{3.01},
  \begin{equation}
\mid \int_{I \times I}  K(x,y) \ dxdy \mid   \leq C \vert
I \vert 
\label{A.1}
  \end{equation}
for all dyadic intervals $I$ (this is the \lq\lq weak cancellation
condition", see \cite{MEYER2}), then \rf{3.02} holds for all $i,l$.

If $T$ is a pseudo-differential operator with symbol
$\sigma(x,\xi)$ defined by the formula
\beq{v101}
T(f) (x) = \sigma(x,D) f = \int e^{i x \xi} \ \sigma(x, \xi)
\hat{f}(\xi) \ d\xi = \int K(x,y) f(y) \ dy,
\eeq
where $K$ is the distributional kernel of $T$, then assuming that the symbols
$\sigma$ of $T$ and $\sigma^* $ of $T^*$ satisfy the standard conditions
\beq{v102}
\mid \partial ^{\alpha}_{\xi} \ \partial ^{\beta}_x \ \sigma(x, \xi)
\mid \leq C_{\alpha , \beta} (1+\mid \xi \mid )
^{\lambda - \alpha + \beta}
\eeq
\beq{v103}
\mid \partial ^{\alpha}_{\xi} \ \partial ^{\beta}_x \ \sigma^*(x, \xi)
\mid \leq C_{\alpha , \beta} (1+\mid \xi \mid )
^{\lambda - \alpha + \beta},
\eeq
we have the inequality
\beq{v104}
|\alpha^j_{i,l}|+|\beta^j_{i,l}| +|\gamma^j_{i,l}|
\leq\frac{2^{\lambda \, j } \ C_M}{(1 + | i-l|)^{M+1}},
\eeq
for all integer $i, l$.

Suppose now that we approximate
the operator $T_0$ by the operator $T_0^{B}$ obtained from $T_0$
by setting to zero all coefficients of matrices
$\alpha^j$, $\beta^j$ and $\gamma^j$ outside of bands of
width  $B \geq 2 M$ around their diagonals. We obtain
\beq{v40a}
\parallel T_0^{B} - T_0 \parallel \leq {C \over B^M }  \log_2 N,
\eeq
where $C$ is a constant determined by the kernel $K$ and $ \log_2 N$ is the number of scales in the representation. In most numerical applications, the accuracy $\varepsilon$ of calculations is fixed, and the parameters of the algorithm (in our case, the band width $B$ and order $M$) have to be chosen in such a manner that the desired precision of calculations is achieved. If $M$ is fixed, then  we choose $B$ so that
\beq{v41b}
B \geq \left( \frac{C}{\varepsilon} \log_2 N \right)^{1/M}.
\eeq
In other words, $T_0$ has been approximated to precision $\varepsilon$
with its truncated version, which can be applied to arbitrary
vectors for a cost proportional to 
$N \left( ( C / \varepsilon) \log_2 N \right)^{1/M}$, 
which for all practical purposes does not differ from $N$. 

A more detailed investigation \cite{BCR1}
permits the estimate \rf{v40a} to be replaced with the estimate
\beq{v41c}
\parallel T_0^{B} - T_0 \parallel \leq {C \over B^M },
\eeq
making the application of the operator $T_0$ to an arbitrary vector
with arbitrary fixed accuracy into a procedure of order
$N$. Obtaining this uniform estimate leads to a proof of 

\vspace{.15in}
\noindent
{\bf  Theorem  (G. David, J.L. Journ\'e)}
Suppose that the operator
  \begin{equation}
 T(f)=\int K(x,y) \ f(y) \ dy
  \end{equation}
satisfies the conditions \rf{3.00}, \rf{3.01}, \rf{A.1}.
Then a necessary and
sufficient condition for $T$ to be  bounded  on $L^2$ is
that
  \begin{equation}
\beta (x)=T(1)(x),
  \end{equation}
  \begin{equation}
\gamma(y)=T^*(1)(y)
  \end{equation}
belong to dyadic $B.M.O.$, i.e. satisfy condition 
  \begin{equation}
\sup_{J} \frac{1}{\vert J\vert}\int_J\vert\beta (x)-m_J(\beta )\vert^2dx\leq  C,
\label{A.6}
  \end{equation}
where $J$ is a dyadic interval and 
  \begin{equation}
m_J(\beta )=\frac{1}{\vert J\vert}\int_J \beta(x) dx.
  \end{equation}
 Again we refer to \cite{BCR1} for details.

The compression of operators results in fast algorithms for evaluation of 
operators on functions. We present here one example and refer to \cite{BCR1} for additional examples.

\bigskip
\noindent
{\bf Example 1.}

In this example, we consider the matrix
$$
A_{ij}= \left \{
\ba{ll}
 \frac{1}{i-j} & i\neq j, \\
0 & i=j,
\ea
\right.
$$
and convert it to the non-standard form using wavelets with six vanishing moments. Setting to zero all entries whose absolute values are smaller than $10^{-7}$, we obtain the non-standard form where the non-zero elements are shown in black in Figure~\ref{Fig2}. The results of experiments in applying this sparse matrix to a vector are tabulated in  Table~1. The standard form of the operator $A$ with $N=256$ is depicted in Figure~\ref{Fig4}.

Column 1 of Table~1 contains the number $N$ indicating the size of $N \times N$ matrix $A_{ij}$,  columns 2, 3 contain CPU times $T_s$, $T_w$ required by the standard order $O(N^2)$ and the fast $O(N)$ schemes to multiply a  vector by the matrix, and column 4 contains the CPU $T_d$ time used to produce the non-standard form of the operator. Columns $ 5, 6$ contain the $L_2$ and $L_{\infty}$ errors of the direct calculation, and columns $7, 8$
contain the same information for the result obtained by computing in the wavelet system of coordinates.  Finally, the last column contains
the compression coefficients $C_{comp}$, defined by the ratio of $N^2$ to the number of non-zero elements in the non-standard form of of the matrix.

\vskip .25in

\begin{center}
\begin{footnotesize}
\begin{tabular}{|r|c|c|c|l|l|l|l|c|}
\hline 
 &  \multicolumn{3}{|c|}{}   & \multicolumn{2}{c}{} &
\multicolumn{2}{|c|}{} &  \\
Input &  \multicolumn{3}{|c|}{Time}   & \multicolumn{2}{c}{Error of Single Precision } &
\multicolumn{2}{|c|}{Error of FWT } & Compression \\
\multicolumn{1}{|c|}{Size} & \multicolumn{3}{|c|}{}   &
\multicolumn{2}{|c|}{Multiplication} &
\multicolumn{2}{|c|}{Multiplication} & Coefficient \\
\hline
\multicolumn{1}{|c|}{N} & $T_s$ & $T_w$ &  $T_d$ &
\multicolumn{1}{|c|}{$L_2$ - norm} & \multicolumn{1}{|c|}{$L_\infty$ - norm} &
\multicolumn{1}{|c|}{$L_2$ - norm} & \multicolumn{1}{|c|}{$L_\infty$ - norm} & 
$C_{comp}$ \\
\hline &&&&&&&& \\
64   & \   0.12  &  0.16  & \ \ 7.76 & 
$1.26 \cdot 10^{-7}$ & $3.65 \cdot 10^{-7}$ & 
$8.89 \cdot 10^{-8}$ & $1.72 \cdot 10^{-7}$ & \ 1.39 \\ 
\hline &&&&&&&& \\
128  & \  0.48  &  0.38  & \ 32.62 & 
$2.17 \cdot 10^{-7}$ & $8.64 \cdot 10^{-7}$ & 
$1.12 \cdot 10^{-7}$ & $9.94 \cdot 10^{-7}$ & \ 2.22 \\ 
\hline &&&&&&&& \\
256  & \  1.92 & 0.80 & \ 96.44 & 
$2.81 \cdot 10^{-7}$ & $1.12 \cdot 10^{-6}$ &
$1.25 \cdot 10^{-7}$ & $5.30 \cdot 10^{-7}$ & \ 3.93 \\
\hline &&&&&&&& \\
512  & \  7.68 & 1.80 & 252.72 & 
$4.21 \cdot 10^{-7}$ & $1.75 \cdot 10^{-6}$ &
$1.23 \cdot 10^{-7}$ & $5.16 \cdot 10^{-7}$ & \ 7.33 \\
\hline &&&&&&&& \\
1024 &   30.72 & 3.72 & 605.74 & 
$6.64 \cdot 10^{-7}$ & $3.90 \cdot 10^{-6}$
& $1.36 \cdot 10^{-7}$ & $5.04 \cdot 10^{-7}$ & 14.09 \\
\hline
\end{tabular}
\end{footnotesize}
\end{center}

\begin{center}
{\bf Table 1: } Numerical results for Example 1
\end{center}

\Section{The operator $d/dx$ in wavelet bases.}
\label{ddx}

For a number of operators (e.g., differential operators, fractional derivatives, Hilbert and Riesz transforms) we may compute the non-standard form in the wavelet bases by solving a small system of linear algebraic equations \cite{BEY}. As an example, we construct the non-standard form of the operator $d/dx$. The matrix elements $\alpha^j_{il}$, $\beta^j_{il}$, and $\gamma^j_{il}$ of $A_j$, $B_j$, and $\Gamma_j$,  where $i,l,j \in {\bf Z}$ for 
the operator $d/dx$ are easily computed as
\beq{70.1}
\alpha^j_{il} = 2^{-j} \int_{-\infty}^{\infty} \,  \psi (2^{-j} x -i) \,
\psi' (2^{-j} x -l) \, 2^{-j} dx = 2^{-j} \alpha_{i-l},
\eeq
\beq{70.2}
\beta^j_{il} = 2^{-j} \int_{-\infty}^{\infty} \,  \psi (2^{-j} x -i) \,
\varphi' (2^{-j} x -l) \, 2^{-j} dx = 2^{-j} \beta_{i-l},
\eeq
and
\beq{70.3}
\gamma^j_{il} = 2^{-j} \int_{-\infty}^{\infty} \,  \varphi (2^{-j} x -i) \,
\psi' (2^{-j} x -l) \, 2^{-j} dx = 2^{-j} \gamma_{i-l},
\eeq
where 
\beq{70.4}
\alpha_l = \int_{-\infty}^{+\infty} \psi(x-l) \, \frac{d}{dx} \psi (x) \, dx ,
\eeq
\beq{70.5}
\beta_l = \int_{-\infty}^{+\infty} \psi(x-l) \, \frac{d}{dx} \varphi (x) \, dx,
\eeq
and
\beq{70.6}
\gamma_l = \int_{-\infty}^{+\infty} \varphi(x-l) \, \frac{d}{dx} \psi (x) \, dx.
\eeq
Moreover, using \rf{1.001} and \rf{1.017} we have
\beq{70.7}
\alpha_i = 2 \sum_{k=0}^{\LF-1} \, \sum_{k'=0}^{\LF-1} \, g_k \, g_{k'} \,
\derc_{2i+k-k'},
\eeq
\beq{70.8}
\beta_i =  2 \sum_{k=0}^{\LF-1} \, \sum_{k'=0}^{\LF-1} \, g_k \, h_{k'} \,
\derc_{2i+k-k'},
\eeq
and
\beq{70.9}
\gamma_i =  2 \sum_{k=0}^{\LF-1} \, \sum_{k'=0}^{\LF-1} \, h_k \, g_{k'} \,
\derc_{2i+k-k'},
\eeq
where 
\beq{7.1}
\derc_l = \int_{-\infty}^{+\infty} \varphi(x-l) \, 
\frac{d}{dx} \varphi (x) \, dx , \quad l \in {\bf Z}.
\eeq
Therefore, the representation of $d/dx$ is completely determined by 
the coefficients $\derc_l$ in \rf{7.1} or in other words, by the representation
of $d/dx$ on the subspace ${\bf V}_0$. Rewriting \rf{7.1} in terms of ${\hat \varphi} (\xi)$ (see \rf{1.003}), we obtain
\beq{7.102}
\derc_l = \int_{-\infty}^{+\infty} |{\hat \varphi} (\xi)|^2 ( {\rm i} \xi)
{\rm e}^{-{\rm i} l \xi} \, d\xi.
\eeq
Thus, the coefficients $\derc_l$ depend only on the autocorrelation function of the scaling function $\varphi$, rather that the scaling function itself since the integral in \rf{7.102} depends just on $|{\hat \varphi} (\xi)|^2$.
The same holds, in fact, for all convolution operators \cite{BEY}. 

\bigskip

\noindent {\bf Remark}. The autocorrelation function of the scaling function (see \rf{17b.3}) has $2M -1$ vanishing moments and its "zero moment" is equal to one  (see \rf{17b.4} and \rf{17b.5}). This implies that if we consider the representation of the derivative operator on the subspace ${\bf V}_0$ as a finite-difference scheme, such scheme has order $2M$. For integral convolution operators, it implies that the asymptotics is accurate to order $2M$ (see \cite{BEY} and below).

\bigskip

The following proposition \cite{BEY} reduces the computation of
the coefficients  $\derc_l$ to solving a system of linear algebraic
equations.

\noindent 

{\bf 1}. If the integrals in \rf{7.1} or \rf{7.102} exist, then
the coefficients $\derc_l$,
$l \in {\bf Z}$ in \rf{7.1} satisfy the following system of linear algebraic
equations
\beq{11.1}
\derc_l = 2  \left[ \derc_{2l} +  \hf \sum_{k=1}^{\LF/2} a_{2k-1}
(\derc_{2l-2k+1} + \derc_{2l+2k-1}) \right] , 
\eeq
and 
\beq{11.3}
\sum_{l} l \,  \derc_{l} = -1,
\eeq
where 
\beq{11.3a}
a_{2k-1} =  2 \sum_{i=0}^{\LF-2k} h_i \, h_{i+2k-1}, \quad k=1,\dots,\LF/2
\eeq
are the autocorrelation coefficients of the filter $H$.

{\bf 2}.
If $M \ge 2$, then equations \rf{11.1} and \rf{11.3} 
have a unique solution with a finite number of non-zero
$\derc_l$, namely, $\derc_l \ne 0$ for
$-\LF+2 \le l \le \LF-2$ and
\beq{11.2}
\derc_l = - \derc_{-l},
\eeq
\bigskip

Solving equations \rf{11.1}, \rf{11.3}, we present 
the results for Daubechies' wavelets with $M=2,3$.
For further examples we refer to \cite{BEY}. 

\noindent {\bf 1}. $M=2$ 
$$
a_1 = \frac{9}{8},   \ \ \ \
a_3 = -\frac{1}{8},
$$
and
$$
\derc_1 = -\frac{2}{3},   \ \ \ \
\derc_2 = \frac{1}{12},   \ \ \ \
$$
We note, that the coefficients $(-1/12,2/3,0,-2/3,1/12)$ of this example
can be found in many books on numerical analysis as a choice of
coefficients for numerical differentiation.

\indent \hrule width5cm height1pt 
 
\bigskip

\noindent {\bf 2}. $M=3$
$$
a_1 = \frac{75}{64},   \ \ \ \
a_3 = -\frac{25}{128},  \ \ \ \
a_5 = \frac{3}{128},
$$
and
$$
\derc_1 = -\frac{272}{365},   \ \ \ \
\derc_2 = \frac{53}{365},   \ \ \ \
\derc_3 = -\frac{16}{1095},   \ \ \ \
\derc_4 = -\frac{1}{2920}.
$$

\begin{figure}
\epsfxsize = 300pt
\centerline{\epsffile{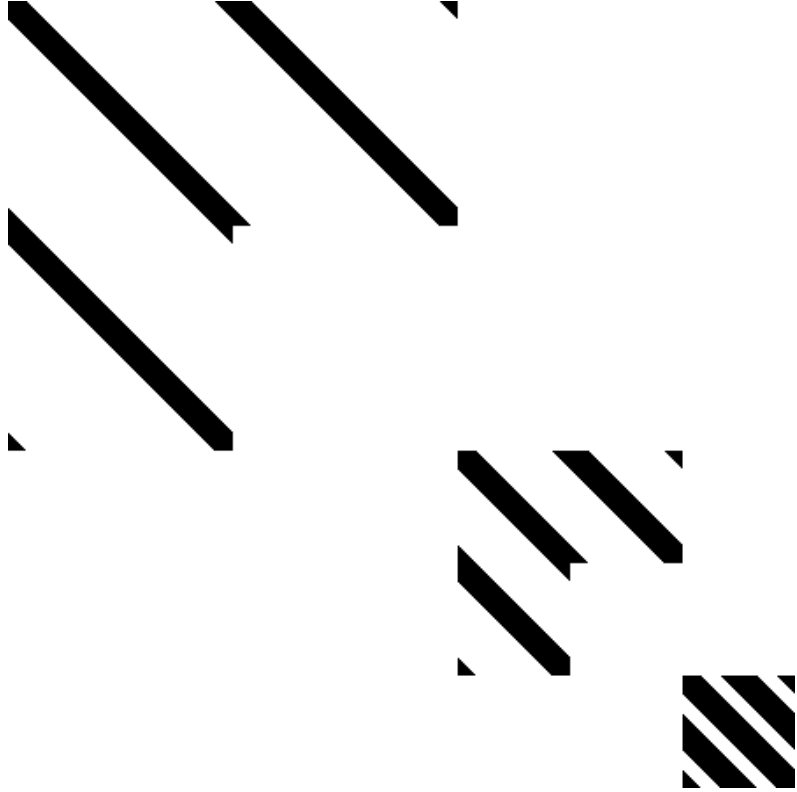}}
\caption{Non-zero entries of a derivative operator in the non-standard form}
\label{FigNSFD2}
\end{figure}

\begin{figure}
\epsfxsize = 225pt
\centerline{\epsffile{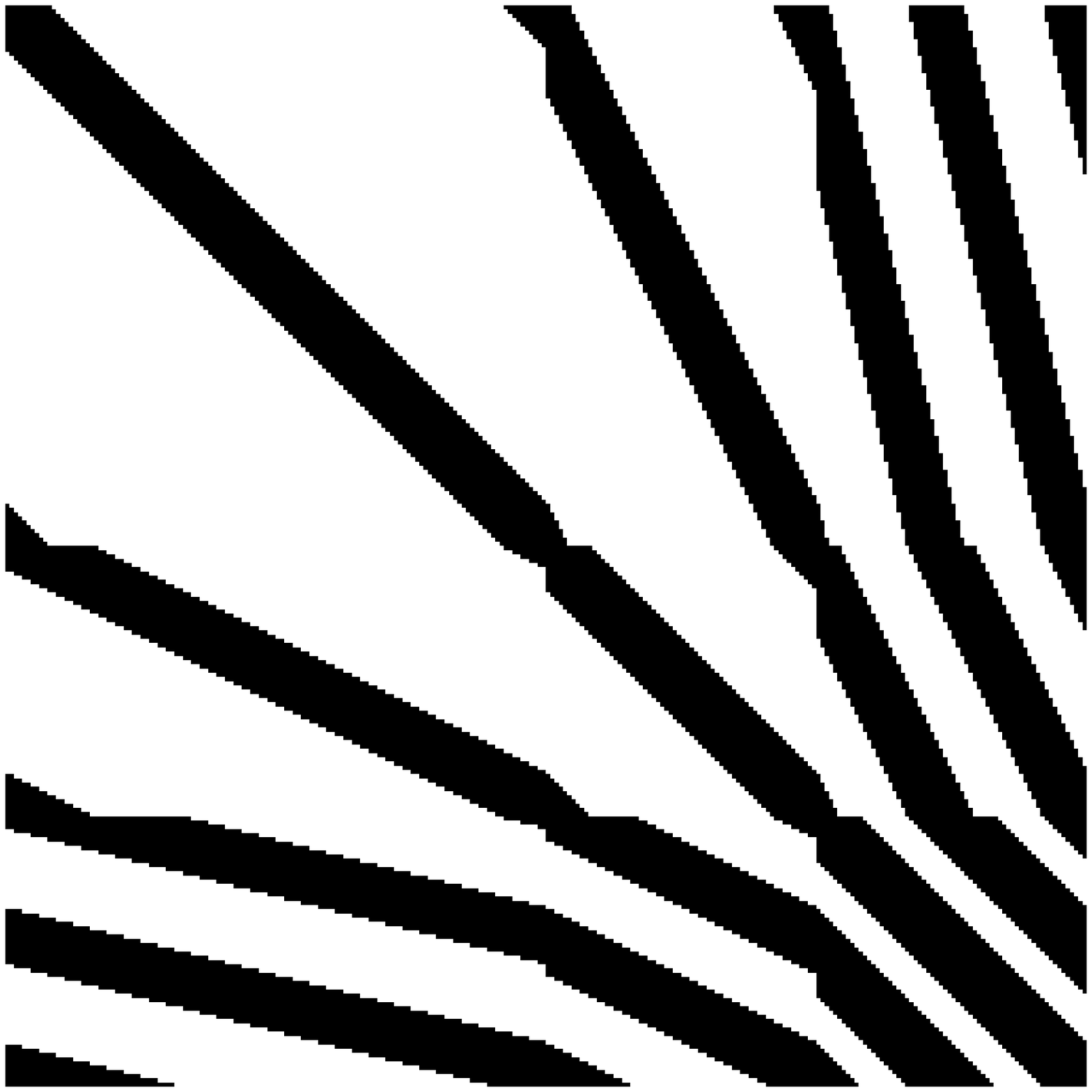}}
\caption{Non-zero entries of a derivative operator in the standard form}
\label{FigSFD2}
\end{figure}

The structure of non-standard and standard forms of derivative operators
is illustrated in Figures~\ref{FigNSFD2}~and~\ref{FigSFD2}.

For the coefficients $\derc_l^{(n)}$ of $d^n/dx^n$, $n > 1$, 
the system of linear algebraic equations is similar to that for the coefficients of $d/dx$. This system (and \rf{11.1}) may be written in terms of 
\beq{15.05}
{\hat \derc}( \xi) 
= \sum_{l} \derc_l^{(n)} {\rm e}^{{\rm i} l \xi},
\eeq
as 
\beq{15.17}
{\hat \derc}(\xi) =
2^{n} 
\left( \, 
| m_0 (\xi /2) |^2  \, {\hat \derc} (\xi/2) + 
| m_0 (\xi/2+ \pi) |^2 \, {\hat \derc} (\xi/2+ \pi)  
\right),
\eeq
where $m_0$ is the $2 \pi$-periodic function in \rf{1.004}.
Considering the operator $M_0$ on $2\pi$-periodic functions
\beq{15.171}
(M_0 f) (\xi) =
| m_0 (\xi /2) |^2 \, f (\xi /2) + 
| m_0 (\xi /2 + \pi) |^2 \, f (\xi /2 + \pi), 
\eeq
we rewrite \rf{15.17} as 
\beq{15.172}
M_0 {\hat \derc} = 2^{-n} {\hat \derc},
\eeq
so that ${\hat \derc}$ is an eigenvector of the
operator $M_0$ corresponding to the eigenvalue $2^{-n}$.
Thus, finding the representation of the derivatives
in the wavelet basis is equivalent to finding trigonometric
polynomial solutions of \rf{15.172} and vice versa \cite{BEY}.

An important property of the wavelet representation of the (periodized) derivative operators (and, in general, pseudodifferential operators with homogeneous symbols) is that these operators have an explicit diagonal preconditioner in wavelet bases.

\begin{figure}
\epsfxsize = 250pt
\centerline{\epsffile{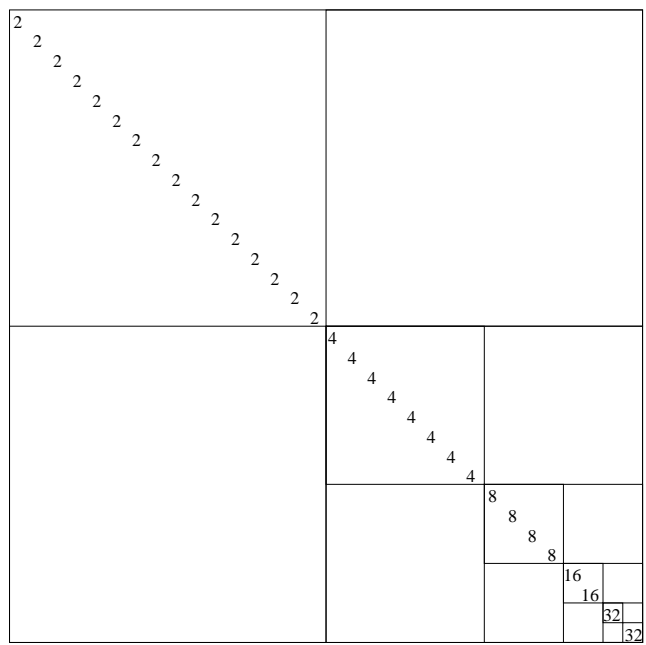}}
\caption{ An example ($n=5$) of the diagonal martix ${\bf P}$ used to rescale the matrix  of the periodized second derivative ${\bf D}_2^w$ in the wavelet system of coordinates.}
\label{FigP}
\end{figure}

We present here two tables illustrating such preconditioning applied
to the standard form of the second derivative.
In the following examples the standard form of periodized
second derivative ${\bf D}_2$ of size $N \times N$, where $N=2^n$, is preconditioned in the wavelet basis by the diagonal matrix ${\bf P}$,
$$
{\bf D}_2^p = {\bf P} {\bf D}_2^w {\bf P},
$$
where 
$P_{il} = \delta_{il} 2^j$, $1 \le j \le n$, and where  
$j$ is chosen depending on $i,l$ so that $N-N/2^{j-1}+1 \le i,l \le N - N/2^{j}$, and $P_{NN}=2^n$.
The matrix ${\bf P}$ is illustrated in Figure~\ref{FigP}.

The following tables compare the original condition number $\kappa$ of ${\bf D}_2^w$ and $\kappa_p$ of ${\bf D}_2^p$.

\bigskip

\begin{center}
\begin{footnotesize}
\begin{tabular}{|c|cc|cc|}
\hline
 &  & &  &  \\
N  & & $\kappa$ &  &  $\kappa_p$ \\
\hline
&  & & &\\
$\ \ 64$     &  &  0.14545E+04             &  & 0.10792E+02  \\
\hline
&  & & &\\
$\ 128$      &  &  0.58181E+04             &  & 0.11511E+02 \\
\hline
& & & &\\
$\ 256$      &  &  0.23272E+05             &  & 0.12091E+02  \\
\hline
& & & &\\
$\ 512$      &  &  0.93089E+05             &  & 0.12604E+02  \\
\hline
& & & &\\
$1024$       &  &  0.37236E+06             &  & 0.13045E+02 \\
\hline
\end{tabular}
\end{footnotesize}

\bigskip

{\bf Table 2.} 

Condition numbers of the matrix of periodized second derivative (with and without preconditioning) in the basis of Daubechies' wavelets with three vanishing moments $M=3$.
\end{center}

\bigskip

\begin{center}
\begin{footnotesize}
\begin{tabular}{|c|cc|cc|}
\hline
 &  & &  &  \\
N  & & $\kappa$ &  & $\kappa_p$ \\
\hline
&  & & &\\
$\ \ 64$     &  &    0.10472E+04           &  & 0.43542E+01  \\
\hline
&  & & &\\
$\ 128$      &  &    0.41886E+04           &  & 0.43595E+01 \\
\hline
& & & &\\
$\ 256$      &  &    0.16754E+05           &  & 0.43620E+01 \\
\hline
& & & &\\
$\ 512$      &  &    0.67018E+05           &  & 0.43633E+01 \\
\hline
& & & &\\
$1024$       &  &    0.26807E+06           &  & 0.43640E+01 \\
\hline
\end{tabular}
\end{footnotesize}

\bigskip

{\bf Table 3.} 

Condition numbers of the matrix of periodized second derivative (with and without preconditioning) in the basis of Daubechies' wavelets with six vanishing moments $M=6$.
\end{center}

\bigskip

\noindent {\bf Fractional derivatives}

First, let us consider convolution operator $T$ and the infinite matrix
$t_{i-l}^{(j-1)}$, $i,l \in {\bf Z}$,  representing $P_{j-1} T P_{j-1}$ on the subspace ${\bf V}_{j-1}$. To compute the representation of $P_j T P_j$, we have (see e.g., formula (3.26) of \cite{BCR1})  
\beq{17.1}
t_{l}^{(j)}  = \sum_{k=0}^{\LF-1} \, \sum_{m=0}^{\LF-1} h_k \, h_m \,  
t_{2l+k-m}^{(j-1)}.
\eeq
It easily reduces to
\beq{17.2}
t_{l}^{(j)}  = t_{2l}^{(j-1)}  + \hf \sum_{k=0}^{\LF/2}
a_{2k-1} \,  ( t_{2l-2k+1}^{(j-1)} + t_{2l+2k-1}^{(j-1)} ).
\eeq
where the coefficients $a_{2k-1}$ are given in \rf{11.3a}.

We also have
\beq{17b.1}
t_{l}^{(j)}  = 
\int_{-\infty}^{+\infty} \int_{-\infty}^{+\infty} \,
K (x-y) \, \varphi_{j,0} (y) \,	\varphi_{j,l} (x) \, dx dy,
\eeq
and, by changing the order of integration, we obtain
\beq{17b.2}
t_{l}^{(j)}  = 
2^{j} \int_{-\infty}^{+\infty} \,
K (2^{j}(l-y)) \, \Phi (y) \, dy,
\eeq
where $\Phi$ is the autocorrelation function of the scaling function $\varphi$,
\beq{17b.3}
\Phi (y)  = 
\int_{-\infty}^{+\infty} \,
\varphi (x) \,	\varphi (x-y) \, dx.
\eeq
It is easy to verify (see \cite{BEY}) that
\beq{17b.4}
\int_{-\infty}^{+\infty} \, \Phi (y) dy  = 1,
\eeq
and 
\beq{17b.5}
{\cal M}_{\Phi}^m = 
\int_{-\infty}^{+\infty} \, y^m \,  \Phi (y) dy  = 0, \quad \mbox{for}
\quad 1 \le m \le 2M-1.
\eeq
The vanishing moments of the autocorrelation function $\Phi$ allow us to compute the elements of the matrix $t_{l}^{(j)}$ for large $l$ and sufficiently fine scales $j \le 0$. Expanding the kernel $K$ in the Taylor series, we obtain from \rf{17b.2}
\beq{17b.6}
t_{l}^{(j)}  = 2^{j} K (2^{j}l) +
\frac{(-1)^{2M} 2^{(2M+1)j}}{(2M)!}  \int_{-\infty}^{+\infty} \,
K^{(2M)} (2^{j}(l-{\tilde y})) \, \Phi (y) \, dy,
\eeq
where ${\tilde y} ={\tilde y}(y,l)$ and $K^{(2M)}$ denotes the $(2M)$th derivative of $K$. The decay of $K^{(2M)} (2^{j}(l-{\tilde y}))$ for large $l$
is faster than that of the original kernel (see \rf{3.00} and \rf{3.01} with an appropriate choice of $M$) and \rf{17b.6} implies a one-point quadrature formula $t_{l}^{(j)}  \approx 2^{j} K (2^{j}l)$ for large $l$ and sufficiently fine scales $j \le 0$.

Computing representations of convolution operators simplifies further if the symbol of the operator is homogeneous of some degree. Let us illustrate this using example of fractional derivatives. Defining fractional derivatives as
\beq{17c.1}
\left(  {\partial }_x^{\alpha} f \right) (x)  = \int_{-\infty}^{+\infty}
\frac{(x-y)^{-\alpha-1}_{+}}{\Gamma(-\alpha)} \, f(y) dy, 
\eeq
where  we consider  $\alpha \ne 1,2 \dots $. If $\alpha < 0$, then \rf{17c.1}
defines fractional anti-derivatives.

The representation of ${\partial }_x^{\alpha}$ on ${\bf V}_{0}$ is determined by the coefficients
\beq{17c.2}
\derc_l =  \int_{-\infty}^{+\infty}  \varphi (x-l) \,
\left(  {\partial }_x^{\alpha} \varphi \right) (x) \, dx, \quad l \in {\bf Z},
\eeq
provided that this integral exists.

The non-standard form
${\partial }_x^{\alpha}=\{ A_j, B_j, \Gamma_j \}_{j \in {\bf Z}}$ is computed via 
$A_j=2^{-\alpha j} A_0$, $B_j=2^{-\alpha j} B_0$,
and $\Gamma_j= 2^{-\alpha j} \Gamma_0$, 
where matrix elements $\alpha_{i-l}$, $\beta_{i-l}$, and
$\gamma_{i-l}$ of $A_0$, $B_0$,
and $\Gamma_0$ are obtained from the coefficients $\derc_l$,
\beq{17c.4a}
\alpha_i =  2^{\alpha} \sum_{k=0}^{\LF-1} \, \sum_{k'=0}^{\LF-1} \, g_k \, g_{k'} \,
\derc_{2i+k-k'},
\eeq
\beq{17c.4b}
\beta_i =  2^{\alpha} \sum_{k=0}^{\LF-1} \, \sum_{k'=0}^{\LF-1} \, g_k \, h_{k'} \,
\derc_{2i+k-k'},
\eeq
and
\beq{17c.4c}
\gamma_i =  2^{\alpha} \sum_{k=0}^{\LF-1} \, \sum_{k'=0}^{\LF-1} \, h_k \, g_{k'} \,
\derc_{2i+k-k'}.
\eeq

It easy to verify that the coefficients $\derc_l$ satisfy the following system of linear algebraic equations
\beq{17c.3}
\derc_l =   2^{\alpha} \left[ \derc_{2l} +  \hf \sum_{k=1}^{\LF/2} a_{2k-1}
(\derc_{2l-2k+1} + \derc_{2l+2k-1})  \right] , 
\eeq
where the coefficients $a_{2k-1}$ are given in \rf{11.3a}. 
Using \rf{17b.6}, we obtain the asymptotics of $\derc_l$ for large $l$,
\beqa
\derc_l & = &  
\frac{1}{\Gamma(-\alpha)}
\frac{1}{l^{1+\alpha}} 
+ O(\frac{1}{l^{1+\alpha+2M}}) \quad  \mbox{for} \quad l > 0, \label{17c.4} \\
\derc_l & = &  0
\quad \quad \quad \quad \quad \quad \quad 
\quad \quad \quad \quad \quad \quad \quad 
\mbox{for} \quad l < 0. \label{17c.5}
\eeqa

\noindent {\bf Example.}

We compute the coefficients $\derc_l$
of the fractional derivative with $\alpha=0.5$ for Daubechies' wavelets
with six vanishing moments with accuracy $ 10^{-7}$. 
The coefficients for $\derc_l$, $l > 14$ or $l < -7$ are obtained using asymptotics 
\beqa
\derc_l & = &  
- \frac{1}{2\sqrt{\pi}} \frac{1}{l^{1+\hff}}
+ O(\frac{1}{l^{13+\hff}}) \quad  \mbox{for} \quad l > 0,  \\
\derc_l & = &  0
\quad \quad \quad \quad \quad \quad \quad 
\quad \quad \quad \quad 
\mbox{for} \quad l < 0. 
\eeqa

\begin{center}
\begin{footnotesize}
\begin{tabular}{|c|c|c|c|}
\hline
&Coefficients  &  &  Coefficients  \\
$l$ & $\derc_l$  & $l$ &  $\derc_l$   \\
 & & &\\
\hline
-7 &     -2.82831017E-06      &  4 &  -2.77955293E-02        
\\
\hline
-6 &     -1.68623867E-06      &  5 & -2.61324170E-02    
\\
\hline
-5 &    4.45847796E-04        &   6 & -1.91718816E-02  
\\
\hline
-4 &     -4.34633415E-03     &  7 & -1.52272841E-02    
\\
\hline
-3 &    2.28821728E-02        &  8 &   -1.24667403E-02  
\\
\hline
-2 &     -8.49883759E-02      &  9 &    -1.04479500E-02  
\\
\hline
-1 &    0.27799963          &     10 &     -8.92061945E-03    
\\
\hline
0 &  0.84681966           &   11 &      -7.73225246E-03    
\\
\hline
1 &    -0.69847577             &   12 &    -6.78614593E-03    
\\
\hline
2 & 2.36400139E-02         &   13 &    -6.01838599E-03  
\\
\hline
3 &    -8.97463780E-02         &   14 &    -5.38521459E-03   
\\
\hline
\end{tabular}
\end{footnotesize}

\bigskip

{\bf Table 5.} 
The coefficients $\{ \derc_l \}_{l}$,
$l=-7,\dots,14$ of the fractional derivative $\alpha=0.5$ for Daubechies' wavelet
with six vanishing moments. 
\end{center}

\Section{Multiplication of matrices and fast iterative construction of the generalized inverse}
\label{Mult}

The standard and non-standard forms may be multiplied in fast manner if the matrices represent Calder\'on-Zygmund or pseudo-differential operators.
Multiplication of matrices in the standard form is a straightforward algorithm \cite{BCRREV}, \cite{ABCR} and requires at most $O(N \log^2 N)$ operations. The algorithm for the multiplication of matrices in the non-standard form has been outlined in \cite{BEY-INRIA} and requires $O(N)$ operations. This is a significant improvement over $O(N^3)$ operations for dense matrices which arise in the ordinary discretization of the operators from these classes.

Fast multiplication algorithms  give a second life to a great number of iterative algorithms. Indeed, powers of matrices maybe computed and so are other functions of matrices. Let us consider an iterative construction of the generalized inverse. In order to construct the generalized inverse  $A^\dagger$ of the matrix $A$,  we use the following result \cite{SCHULZ}:

\vspace{7 mm}

{\it
Let $\sigma_1$ be the largest singular value of the $m \times n$ matrix
$A$. Consider the sequence of matrices $X_k$
\beq{2.i01}
X_{k+1}=2 X_k- X_k A X_k
\eeq
with
\beq{2.i02}
X_0=\alpha A^*,
\eeq
where $A^*$ is the adjoint matrix and  $\alpha$ is chosen so that the largest
eigenvalue of $\alpha A^*A$ is less than one.
Then the sequence $X_k$ converges to the generalized inverse $A^\dagger$.
}

\vspace{7 mm}

Combining this iteration with fast multiplication algorithms, we obtain an algorithm for constructing the generalized inverse in at most  $O(N \log^2 N \log R)$ operations, where $R$ is the condition number of the matrix. 
(By the condition number we understand the ratio of the largest singular value to the smallest singular value above the threshold of accuracy). 

The details of this algorithm (in the context of computing in wavelet bases)
will be described in \cite{BCR2}. We note that throughout the iteration \rf{2.i01}, it is  necessary to maintain the \lq\lq finger" band structure of the standard form  of matrices $X_k$.   Hence, the standard form of both the operator and its generalized inverse must admit such structure.  We note that the pseudo-differential operators satisfy this condition.

The following table contains timings and accuracy comparison of the constructionof the generalized inverse
via the singular value decomposition (SVD), which is $O(N^3)$ procedure, 
and via the iteration \rf{2.i01}-\rf{2.i02} in the wavelet basis
using Fast Wavelet Transform (FWT).
The computations were performed on Sun Sparc workstation and
we used a routine from LINPACK for computing the singular value decomposition.
For tests we used the following full rank matrix
$$
A_{ij} = \left \{
\ba{ll}
\frac{1}{i-j} & i\neq j \\ \\
1 & i=j 
\ea
, \right.
$$
where $i,j=1,\dots,N$. The accuracy theshold was set to $10^{-4}$, i.e.,
entries of $X_k$ below $10^{-4}$ were systematically removed after each iteration.

\bigskip

\begin{center}
\begin{footnotesize}
\begin{tabular}{cccc}
Size  $N \times N $ & SVD            &  FWT Generalized Inverse & $L_2$-Error   \\ \\
$128 \times 128$   & 20.27  sec.       &  25.89 sec.   &   $3.1 \cdot 10^{-4}$  \\ \\ 
$256 \times 256$   & 144.43  sec.      &  77.98 sec.   &   $3.42 \cdot 10^{-4}$ \\ \\
$512 \times 512$   & 1,155 sec. (est.) &  242.84  sec. &   $6.0 \cdot 10^{-4}$  \\ \\
$1024 \times 1024$ & 9,244 sec. (est.) &  657.09  sec. &   $7.7 \cdot 10^{-4}$  \\ \\
\dots          & \dots           & \dots                    & \dots \\ \\
$2^{15} \times 2^{15}$ &  9.6 years (est.) & 1 day (est.)     &       \\ \\
\end{tabular}
\end{footnotesize}
\end{center}

We note that the iteration in \rf{2.i01} also allows us to compute the projector on the null space (see \cite{BCRREV} for this and several other examples). 

The algorithm for the exponential is based on the identity
\beq{5.01}
\exp(A)=\left[ \exp(2^{-L}A) \right]^{2^{L}}.
\eeq
First, $\exp(2^{-L}A)$ is computed by, for example,  using the Taylor series.
The number $L$ is chosen so that the largest singular value
of $2^{-L}A$ is less than one. At the second stage of the algorithm
the matrix $2^{-L}A$ is squared $L$ times to obtain the result.
Similarly, sine and cosine of a matrix can be computed using the elementary
double-angle formulas. 
Unlike the algorithm for the generalized inverse, this algorithm is not self-correcting. Thus, it is necessary to maintain sufficient accuracy initially so as to obtain the desired accuracy after all the mutiplications have been performed.

\begin{figure}
\epsfxsize = 300pt
\centerline{\epsffile{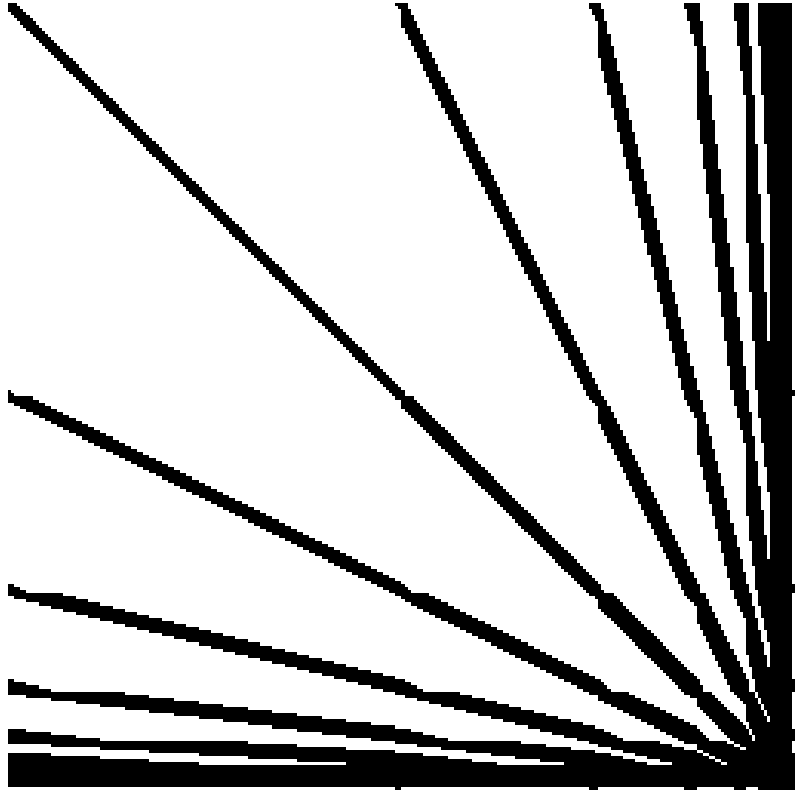}}
\caption{ Matrix ${\bf D_b}^{-1}$ computed via iterative algorithm of this Section with diagonal rescaling. Entries with the absolute value greater than $10^{-8}$ are shown black.}
\label{FigBVPi}
\end{figure}

Finally, as an example, let us consider the matrix 
\beq{eq7.1}
{\bf D_b}= 
\left(
\begin{array}{ccccccc}
-2 & 1 &     0  & \cdots & 0 & 0  & 0 \\
1 & -2 & 1 &    \cdots & 0 & 0  & 0 \\
\cdots &  \cdots   & \cdots & \cdots  &  \cdots & \cdots & \cdots \\
0  & 0 & 0  &  \cdots & 1 & -2 & 1 \\
0   & 0 & 0   & \cdots   & 0  &  1 &   -2 
\end{array} 
\right).
\eeq
which arises in the finite-difference formulation of the two-point boundary value problem. We note that the inverse of this matrix is sparse in the wavelet basis. As an illustration, in Figure~\ref{FigBVPi} we display the inverse matrix ${\bf D_b}^{-1}$ computed starting with matrix ${\bf D_b}$. Using the diagonal preconditioning (see Figure~\ref{FigP}), this computation involves only well-conditioned matrices \cite{BEYBVP}. 

\bibliography{/beylkin/biblio/biblio} 
\end{document}